\documentclass{article}

\usepackage{graphicx}
\usepackage{rotating} 
\usepackage{fullpage,amsmath,amssymb,latexsym}
\usepackage{multirow}
\usepackage{natbib}

\begin{document}

\title{The Effects of Metallicity and Grain Growth and Settling on the Early Evolution of Gaseous Protoplanets}

\author{R. Helled$^1$ and P. Bodenheimer$^2$\\
$^1$Department of Earth and Space Sciences,\\
University of California, Los Angeles, CA 90095 1567, USA\\
$^2$Department of Astronomy and Astrophysics,\\
UCO/ Lick Observatory, University of California, Santa Cruz, \\
CA 95064, USA\\
rhelled@ucla.edu (R. H.); peter@ucolick.org(P. B.)}

\date{}
\maketitle %\vskip 5cm \noindent

\begin{abstract}
Giant protoplanets formed by gravitational instability in the outer
regions of circumstellar disks go through an early phase of
quasi-static contraction during which radii are large ($\sim 1$
AU) and internal temperatures are low ($< 2000$ K). The main source of
opacity in these objects is dust grains. We investigate two problems
involving the effect of opacity on the evolution of isolated, non-accreting planets of
3, 5, and 7 M$_J$. First, we pick three different overall metallicities
for the planet and simply scale the opacity accordingly. We show that
 higher metallicity results in slower contraction as a result of
higher opacity. It is found that the pre-collapse time scale is
proportional to the metallicity. In this scenario, survival of
giant planets formed by gravitational instability is predicted to
be more likely around low-metallicity stars, since they evolve to
the point of collapse to small size on shorter time scales. But
metal-rich planets, as a result of longer contraction times, have
the best opportunity to capture planetesimals and form heavy-element
cores. Second, we investigate the effects of opacity reduction as a
result of grain growth and settling, for the same three planetary
masses and for three different values of overall metallicity. When
these processes are included, the pre-collapse time scale is
found to be of order 1000 years for the three masses, significantly
shorter than the time scale calculated without these effects. In this
case the time scale is found to be relatively insensitive to planetary
mass and composition. However, the effects of planetary rotation and accretion of gas and dust, which could increase the timescale, are not included in the calculation. The short time scale we find would preclude metal
enrichment by planetesimal capture, as well as  heavy-element core formation, 
over a large range of planetary masses and metallicities.
\end{abstract}

\section{Introduction}
One suggested mechanism for giant planet formation is gravitational (disk) instability 
in which giant planets are formed by fragmentation in the protoplanetary disk (e.g., Cameron, 1978; Boss, 1997). 
The standard model for giant planet formation , however, is core accretion, 
 in which the heavy-element core forms first by accretion of planetesimals and later captures
 a gaseous envelope (e.g., Pollack et al., 1996; Hubickyj et al.,
 2005). While solar-system-like giant planets are most likely to be formed by the core
 accretion mechanism, the disk instability model may be responsible for the formation of 
massive gaseous planets at very large radial distances (e.g., Rafikov, 2007; Cai et al., 2009).
 Giant planets such as those       detected recently by direct imaging (Kalas et al., 2008;
 Marois et al., 2008) cannot easily be explained with the standard core accretion model due
 to the extremely long  core formation timescales at large distances from the star
 (e.g., Dodson-Robinson et al., 2009). On the other hand, theoretical work suggests that
 gravitational instabilities are likely to occur at radial distances beyond 50 or even 100 AU
 (Boley, 2009; Rafikov, 2009), supporting the idea that massive gaseous planets on wide
 orbits may form by that mechanism. \par 

Generally speaking, the main weakness of the gravitational instability model for 
giant planet formation is the uncertainty regarding whether clumps that are formed in the
 disk can actually survive and evolve to become gravitationally bound planets.  While many of
 the numerical simulations have found   disks that are gravitationally unstable under
 given conditions and have detected  formation of clumps, only in a few cases have the objects
been found  to survive on  longer timescales (see review by Durisen et al., 2007). 
Much work has been devoted to the question of whether clumps can cool fast enough to survive 
tidal disruption (e.g., Boss, 2008; Rafikov, 2007). Currently, due to limited resolution
 and computer power, three-dimensional hydrodynamic  models cannot closely follow the evolution
 of the forming clumps long enough to verify whether these objects can indeed survive.  \par

The evolution of protoplanets formed by gravitational instabilities includes three main stages. 
% with low internal densities and temperatures. 
During the first stage, the newly-formed planetary object is cold and extended with a 
radius of a few thousand times Jupiter's present radius (R$_J$), with hydrogen being in
 molecular form (H$_2$). The clump contracts quasi-statically on time scales of $10^4-10^6$ years,
depending on mass, and as it contracts its
 internal temperatures increase.   Once a central temperature of $\sim$ 2000 K is reached,
 the molecular hydrogen dissociates, and a dynamical collapse of the entire protoplanet occurs 
(the second stage). The extended phase is known as the 'pre-collapse stage' (Decampli and Cameron,
 1979), and during that phase the object is at most risk to be destroyed by tidal disruption
 and disk interactions. After the dynamical collapse the planet becomes compact and dense, with
 the radius being a few times R$_J$. During this third stage it is therefore less likely to 
be disrupted, although the planetary object still has the danger of falling into its parent star
 due to inward migration. The protoplanet then continues to cool and contract on a much longer
 timescale ($10^9$ years). \par

During the extended state the protoplanet is most vulnerable to disk torques and tidal encounters, and therefore
 its survival is questionable. Shortening the pre-collapse timescale may reduce the risk of
 disruption since the transition to a "point-like" (gravitationally bound) protoplanet occurs faster. The time scale of the pre-collapse stage is also important with respect to the
 final planetary composition and structure. During the extended phase the protoplanet can
 capture heavy elements in the form of planetesimals and enrich the interior with these
 elements (e.g., Helled et al., 2006). Cores can be formed in these objects by settling of heavy
 elements to the planetary center as long as the internal temperatures are low (e.g.,
 Helled et al., 2008).  As a result, a longer extended phase could lead to further 
enrichment with heavy elements and would provide more time for forming cores. The more massive
the protoplanet, the shorter the  pre-collapse timescale (Helled and Bodenheimer, 2010, 
hereafter paper I). This may imply that in the disk instability scenario more massive 
planetary objects are likely to survive while lower mass planets (if they survive) will tend
 to be enriched with heavy elements and contain cores. 

An important parameter that controls the contraction timescale is the planetary opacity. 
Lower opacity leads to faster cooling and therefore shortens the pre-collapse phase 
(e.g., Pollack et al., 1996). Below we investigate the effects of opacity reduction due
 to different metallicities (section 2) and grain growth and sedimentation (section 3) on
 the planetary evolution. The results are discussed in section 4 and the 
conclusions summarized in section 5. 

\section{Planetary Metallicity}

The question of whether stellar (and disk) metallicity affects the frequency of giant planet
 formation  by gravitational instabilities has been investigated by different authors. 
Boss (2002) varied the assumed opacity in the models by factors of 10 and 0.1, and found little difference in the results. As a result, Boss (2002)
suggested that gravitational collapse is insensitive to the system's metallicity. 
Cai et al. (2006a,b) investigated the sensitivity to opacity in their models, but found that changing the opacity by factors of 0.25 
to 2 led to enhanced fragmentation when the opacity was lowered. 
Meru and Bate (2010) also find, from three-dimensional radiation-hydrodynamic simulations, that low disk opacity tends
to allow the disk to cool faster and promotes fragmentation (Gammie, 2001); therefore 
lower metallicity is more likely to result in fragmentation. 
Finally, Mayer et al. (2007) found enhanced fragmentation when the mean molecular 
weight $\mu$ was increased, but required rather large increases in $\mu$ (with no change of opacity), from 2.4 to 2.7, to make a significant effect, and concluded that increasing metallicity would result
 in more fragmentation. 
All these groups reached different conclusions in some sense, though it should be noticed that the investigations were somewhat different. 
At present, the topic of fragmentation's sensitivity
 to metallicity is still unsolved.\par
 %Boss (2002) and Mayer et al. (2007). 
 
 Thus, metallicity can also be significant in the context of the protoplanets' evolution.
 Planetary metallicity (opacity) has a fundamental role in governing the cooling timescale of the
 planetary object and therefore the protoplanet's contraction. As a result, metallicity has a
 direct impact on the pre-collapse evolution of the newly-formed planets, and
 possibly their survival. \par  

In this section we present the pre-collapse evolution (ending when central temperatures 
reach $\sim$ 2000 K) for different planetary masses assuming different metallicities. 
Extrasolar planets have been detected around stars with very different metallicities, with
 the metallicity of the parent stars ranging between [Fe/H]$\sim$ -1 and [Fe/H]$\sim$ +0.5.
  Here,  to model protoplanets with different metallicities, we simply multiply the 
grain opacity, calculated with an interstellar size distribution and solar  abundances, 
by the corresponding metallicity. The Rosseland mean opacities
 are obtained from Pollack et al. (1985) and Alexander and Ferguson (1994). Thus, to model the 
evolution of a planetary object with [Fe/H] = -0.477, we multiply the solar grain opacity
 by a factor of 1/3 everywhere.  The planets are assumed to have solar abundances of
hydrogen and helium; the equation of state is essentially an ideal gas of neutral He and
molecular H. The standard stellar structure equations are solved (Bodenheimer et al. 1980)
, with surface 
boundary conditions appropriate for a gray photosphere 
\begin{equation}
  L = 4 \pi R^2 \sigma_B T_{\rm eff}^4~~~~{\rm and} ~~\kappa_R P = \frac{2}{3}g
\end{equation}
where $L$ is the total luminosity, $R$ is the outer radius, $\sigma_B$ is the
Stefan-Boltzmann constant, $T_{\rm eff}$ is the surface temperature,  and $g$, $P$, 
and $\kappa_R$ are, respectively, the
acceleration of gravity, the pressure, and the Rosseland mean grain opacity at the surface.
\par 

The initial radii of the protoplanets were chosen to fall inside the tidal radius at 20 AU for a 3 M$_J$ protoplanet. Since all of the planets have an initial radius of $\sim$ 2 AU they all fall within the Hill radius at distances larger than 20 AU. At smaller radial distances, planets could be tidally disrupted.  Our initial radii are in good agreement with 3D numerical simulations of clump formation in gravitationally unstable disks (Boley et al., 2010).  
\par

Figures 1-3 present the evolutionary tracks for three planetary masses: 3, 5 and 7 M$_J$, 
with solar, 3$\times$solar and, solar/3 compositions ( [Fe/H] = 0, +0.477, and -0.477, 
respectively). More massive protoplanets have shorter pre-collapse evolutions. Protoplanets
 with similar masses but higher metallicities have longer evolutionary paths due to the higher
 opacity that results in slower energy loss. Objects with low metallicities on the other
 hand, have short pre-collapse timescales due to their ability to radiate energy efficiently
 (lower opacities), and therefore contract on shorter timescales. The results suggest
 that a reduction of the opacity everywhere by a factor of $\sim$ 3 compared to solar
 abundance results in a pre-collapse phase shorter by a similar factor. In a similar way,
 increasing the metallicity by a factor of three results in an evolutionary timescale
 that is longer by about a factor of three. 
During the pre-collapse stage the protoplanets behave nearly as polytropes, and the radius is
 scaled inversely to the central temperature. Since the opacity for metal-rich planets is higher,
 for a similar radius the luminosity (as well as effective temperature) will be lower for
 metal-rich objects. Metal-poor protoplanets have higher luminosities and effective 
temperatures for a similar radius. \par

A planetary object of three Jupiter masses can contract for $\sim 2\times 10^5$ years before it
 reaches central temperatures of 2000 K, if its composition is three times solar. Such a slow
 contraction could support the possibility of core formation and additional enrichment in
 heavy elements via planetesimal capture (e.g., Helled et al., 2006, 2008). These processes,
 which can take place during the precollapse evolution, can affect the final compositions and
 internal structures of the planetary objects. 
The analysis suggests that the metallicity (opacity) of the protoplanet (and therefore stellar
 metallicity) has an important impact on its evolution. Metal-rich objects have a better 
opportunity to capture additional solids and form heavy-element cores than metal-poor objects.
On the other hand,  the result may lead to the conclusion that survival of giant planets
 formed via gravitational instability is more likely to occur around low metallicity stars
 due to the shorter pre-collapse timescale, which minimizes the period in which the 
protoplanets are most vulnerable for destruction. However, this conclusion is based on the
 assumption that the grains in the planet remain well mixed throughout the interior and do 
not 
change their size distribution during the evolution.  In fact, grains can grow, and settle,
resulting  in a reduction of  the planetary opacity (e.g., Podolak, 2003; Helled et al.,
 2008; Movshovitz and Podolak, 2008). This possibility is investigated in the next section.

\section{Opacity Reduction Caused by Grain Coagulation and Sedimentation}
As protoplanets contract, grains in their atmospheres can grow, settle, and reduce the planetary
 opacity (Podolak, 2003). The radiative thin outer region of the protoplanet is of primary interest
, since this region essentially controls the cooling of the planet. The evolutionary
calculations described in the previous section show that most of the interior, below the
radiative outer layer, is convective.  
To model grain coagulation and sedimentation and to compute the effect on the planetary opacity
 we use a numerical procedure recently presented by Movshovitz and Podolak (2008), kindly
 provided by N. Movshovitz. A detailed description of the calculation can be found in
  Movshovitz and Podolak (2008) and references therein. Below we briefly discuss the physics behind the numerical procedure. \par

At each level $r$ in the protoplanet at time $t$ the number density of grains with mass between
 $m$ and $m+dm$ is given by $n(m,r,t)dm$.  The evolution of the grain distribution due to
 collisions is described by the Smoluchowski equation 
$$\frac{\partial n(m,r,t)}{\partial t}={\frac 12}\int_0^m\kappa (m^{\prime
},m-m^{\prime })n(m^{\prime },r,t)n(m-m^{\prime },r,t)dm^{\prime
}$$

\begin{equation}
-n(m,r,t)\int_0^\infty \kappa (m,m^{\prime })n(m^{\prime
},r,t)dm^{\prime } -\nabla \cdot F
\end{equation}
where $\kappa (m,m') = \gamma P(m,m')$ is the collision kernel, giving
the probability that a grain of mass $m$ will collide with and stick to a grain of mass $m'$,
 where $P(m,m')$ is the collision probability. We set the sticking coefficient, $\gamma$, to
 be unity so whenever two grains collide they stick. The first integral on the right hand side 
equals the rate of
formation of grains of mass $m$ by collisions between a grain of
mass $m'$ and a grain of mass $m-m'$. The second integral is the
rate of removal of grains of mass $m$ when such a grain combines
with a grain of any other mass. The term involving $F$ is the transport term.
This can be either via gravitational settling through the gas or via turbulent transport if
 the gas is convective. The convective velocities are derived from the evolutionary model
 using the Mixing Length Recipe. Collisions due to the Brownian motion of the grains are 
included,  as is     the fact that larger grains sediment faster than smaller grains and can 
overtake them. \par

In convective regions, small grains can be carried with the convective eddy while large
 grains would not. The different relative velocities between different sized grains result in a change in the coagulation kernel.  
 The relative velocities in the presence of convection of grains with arbitrary sizes were recently presented by Ormel and Cuzzi (2007). 
The code uses Ormel and Cuzzi's  expressions but also includes the sedimentation speeds. The sedimentation velocity 
changes the boundary between large grains which are unaffected by the convective eddies and small grains that are coupled to the gas. The 
difference in sedimentation speed is added in quadrature to the relative velocity in the presence of convection. The grain distribution in convective regions is calculated
using a simple algorithm that mimics the eddy diffusion approximation. When convection is found to affect the grain motion, i.e. (1) the grain's sedimentation speed is smaller than 
the speed of the eddy and (2) the grain can be entrapped in the largest eddy (whose
size is defined as 1.5$\times$ the pressure scale height), grains are redistributed so that 
the ratio of mass density of grains to the gas' mass density is constant. The grain mass in the convective regions can then be conserved.
 \par

The density of the grains is assumed to be 2.8 g cm$^{-3}$, and the initial grain size $a_0$ is taken to
 be $10^{-5}$ cm. The grains are taken to be composed primarily of silicates.  Evaporation of grains is not included in the calculation. However, evaporation of grains is negligible at the internal temperatures of our models. The initial size is not of significant importance as long as it is relatively 
small. The grains are initially homogeneously mixed within the protoplanet.  The size distribution is divided among bins with radii a$_i$ that are logarithmically spaced in mass. The size of a grain in bin $i$, $a_i$, is taken to be $a_02^{i/3}$. The total number of bins is chosen to be 40 so most of the grains are accommodated and can grow to large sizes. Initially, only the smallest mass bin is populated, and as the grains collide and grow the larger size bins are filled. Since some grains could reach the largest bin and therefore can grow further,
 the code can add an extra bin in which the larger grains are accumulated. At the end of the
 time step the largest grains are returned to the largest bin. This procedure ensures grain mass conservation. \par

The timesteps of the sedimentation code are chosen to be small enough to ensure that by the end of the timestep the change in 
number density in each atmospheric layer, and every size bin does not exceed one percent. The maximum timestep is set to be $10^6$ seconds. More details on the numerical procedure can be found in Movshovitz et al. (2010).   \par

 The mass fraction of grains (1\% for solar composition) is constant for a given metallicity. 
Since the dust-to-gas ratio changes with metallicity, metal-rich protoplanets naturally have
 larger grain number densities. When grains manage to grow large enough they can gain high
 enough sedimentation speeds to settle from the radiative region and reduce the atmospheric
 opacity (see Podolak, 2003). 
At each stage of the evolution, given a planetary   model, the grain opacity $\kappa$ is
 calculated using Mie theory (van de Hulst 1957). The terms for scattering and absorption 
 efficiencies are summarized in Movshovitz and Podolak (2008). The evolution of the grains
is calculated over a given time interval, typically 250 years, for a fixed protoplanetary
structure.  The grain opacity values at
 different atmospheric heights, at the end of the time interval,
 are then entered into the evolutionary model. The planetary 
evolution is then calculated over the same 250-year interval, with the grain opacity
a function of time, interpolated in time between the grain distributions at the beginning
and end of the interval.  At the end of the time interval, the new planetary structure
is used to evolve the grains for another 250 years, and the process is repeated. Again, 
 we consider protoplanets with masses of 3, 5, and 7 M$_J$, and account for three different
 metallicities: solar, solar/3, and 3$\times$solar. 

We find that in all cases the opacity can be reduced substantially, leading to much shorter 
contraction timescales than those found in the previous section.
Figure 4 presents the opacities as a function of depth in the planet
 at different times for 3, 5, and 7 M$_J$ with low (solar/3)
 and high (solar$\times$3) metallicities. 
Grain growth and opacity reduction are most efficient when the number density of grains is
 large, allowing frequent collisions. As a result, opacity reduction is found to be greatest for
 metal-rich and massive protoplanets. 
 
Initially, the protoplanets are found to 
be fully convective with thin radiative outer layers. As the evolution progresses, the grains 
settle out of the radiative layer and the opacity is reduced, causing an increase in 
luminosity. Higher luminosity increases the radiative gradient and promotes convection.
The outer radiative zone gradually disappears and the object becomes fully 
convective, apart from the radiative surface boundary condition. The convection results
in the mixing of grains back up to the surface and the opacity increases again.
This process can be seen in detail in figure 4. For example, 7 Mj with high metallicity starts with a
fairly flat opacity curve, the opacity then decreases with time especially in the surface layers, but at 500 years the planet becomes fully 
convective and the opacity in the outer layers increases.
At still later times (700 years), the outer layers become radiative again and the opacity decreases significantly. 
A similar behavior occurs in some of the other cases.\par

Figure 5 shows the grain number density as a function of grain size for a metal-poor 7 M$_J$ planet at different times, and at three different layers. At time zero only grains with sizes $a_0=10^{-5}$ cm exist. 
Since the planet contracts with time the layers are defined by the percentage of planetary mass instead of an absolute height. The first layer is taken to be the uppermost layer (the surface), at this layer the 
mass integrated outward from the center equals the planetary total mass (m = 100\%). We present two other layers defined by integrated masses of 70\% and 50\% of the total mass.  
The upper panel corresponds to the planetary surface. As can be seen from the figure, at 100 years (black dots) there is already a population of larger-size grains due to grain growth at the surface. 
Grain growth and settling at the surface keep removing small grains as long as the planet's surface is radiative. When the planet becomes fully convective small grains are mixed throughout the entire envelope and are carried back to the surface (red dots). As convection continues more larger grains can be carried to the surface leading to a nearly flat size distribution, although small grains are slightly depleted because they are carried to deeper layers by convection (blue dots). Once the planet's atmosphere becomes radiative again efficient grain growth and sedimentation continue with large grains settling to deeper layers. At deeper regions within the protoplanet grain sedimentation provides a source of larger grains, and in these regions the number density of large grains is actually increasing with time (see middle and bottom panels). When the envelope becomes convective the mixing of small grains leads to an increase in the number density of small sizes deeper in the envelope. The inner regions are less effected by convective mixing, and the size distributions at 600 years and 850 years (not shown) are similar. As time progresses small grains are more depleted due to continuous grain growth and settling. \par 
 
Figures 6-8 present the evolutionary tracks for the three
 planetary masses: 3, 5 and 7 M$_J$, with solar, 3$\times$solar and, solar/3 compositions when 
grain growth and sedimentation are included.  
We find that the 3 M$_J$ protoplanet will collapse after $\sim$ 1,000, 1600, and 2300 years
respectively,  for  metal-rich, solar, and metal-poor compositions.  These timescales are
 significantly shorter compared to the ones ignoring grain coagulation and sedimentation (section 2). 
5 M$_J$ protoplanets have even shorter precollapse timescales, with the shortest found for
 the metal-rich case for which the evolution time is found to be only 700  years. For
 solar and (solar/3) compositions the times are just over 1,500 years. Massive protoplanets 
with masses of 7 M$_J$ are found  to have contraction timescales of 670, 1300, and 1280
  years for metal-rich, solar and metal-poor compositions, respectively. \par

As can be seen from the figures, the shortest evolution is found for the metal-rich
 protoplanets (red curves), which is the opposite trend to that shown in Figures 1--3. 
Although metal-poor objects start with a lower opacity and therefore a higher
luminosity, after about a thousand years the metal-rich protoplanets actually achieve lower
 opacity values. While metal-rich protoplanets have more grains and therefore initially
 have higher opacities, grain growth and settling become very efficient in their atmospheres
 after a relatively short time and the opacity decreases significantly. Metal-poor
 protoplanets on the other hand start with low opacities due to  a lower dust-to-gas 
ratio, but grain growth and sedimentation are less efficient due to the lower number
 densities and less frequent collisions. 
The result, that metal-rich protoplanets evolve faster,  can be understood by using the 
approximation presented in the appendix. \par 

To test the robustness of the results we have redone the computation taking smaller
 time intervals for the opacity recalculation. Figure 9 compares the evolution of a metal-poor and metal-rich 7 M$_J$ protoplanets (solar/3 and 3$\times$solar metallicities, respectively) with time intervals of 250 years (black, red) and 50 years (gray, pink) between opacity calculations. Time intervals shorter than 50 years between opacity calculations would already be significantly smaller than the characteristic coagulation and sedimentation timescales and could cause numerical effects.  
The smaller timesteps allow us to study  in more  detail 
how the opacity reduction affects the mechanism by 
which energy is transferred within the planet. The planetary luminosity is the physical
parameter that is most affected by the opacity. Initially the luminosity
remains nearly constant, as the density in the outer layers is low and there has not
been enough time for significant grain settling. As the evolution progresses and opacity is reduced the luminosity increases. 
Once the protoplanets become fully convective small grains remix to the surface and the luminosity decreases. 
The resulting drop in luminosity then again produces an outer radiative zone. Once this is
present the grain settling results in an increase in luminosity, and the cycle is
repeated. A gradual upward trend of the average luminosity occurs near the end of
the evolution because the average grain size has increased through a significant
part of the mass of the outer envelope of the planet, resulting in overall lower opacity. \par 
 
While the global evolution is unchanged for metal-poor 7 M$_J$, with similar temperatures and radii, there is a noticeable difference in the behavior of luminosities. 
The luminosity in the high-time-resolution case
oscillates around the curve for the low time resolution. The pre-collapse timescale is found to change only by a small fraction, from 1336 years in the low-time-resolution case to 1278 years. \par

Also for 7 M$_J$ with 3$\times$solar metallicity the physical parameter which is most affected by the opacity calculation time-intervals is the luminosity. However, unlike in the low-metal case, the total evolutionary time is found to change from 470 years to 670 years, a relatively large difference. The cause for this difference is related to the effect of opacity on the planetary energy transport mechanism. With too large time-intervals between opacity calculations transitions between convective and radiative envelopes can be skipped, resulting in different evolutionary tracks. The evolution with smaller time-intervals is therefore more accurate. Nevertheless, it should be noted that in both cases the evolution to the onset of collapse is found to be shorter than 1000 years.  
The solar case should be about the same as the metal-poor case due to their very similar evolutions (see figure 8).

\section{Implications on Core Formation and Accretion of Solids}
Accretion of planetesimals is most efficient when the protoplanets are extended and fill
 most or all of their feeding zone; planetesimals are then captured due to gas drag. 
Core formation is possible as long as the central temperatures are low enough to allow 
the existence of solid material that can settle to the center. Since both of these
 processes could take place mostly during the pre-collapse evolution, a change in the
 pre-collapse timescale will significantly impact their efficiency. \par

The results presented in section 2, in which the opacity was assumed to be proportional
 to metallicity, suggest that the pre-collapse timescale is proportional to the 
planetary metallicity, so metal-rich protoplanets would have slower contraction
 and  more time to accrete planetesimals and form cores. In this scenario, 
 giant planets around metal-rich stars, if formed by gravitational instabilities, would 
be enriched with heavier elements compared to their host star, and would have larger cores
than solar-composition planets. However it should be noted that if gravitational
 instabilities occur mostly at very large radial distances, the planetesimal accretion
 efficiency decreases significantly, and the protoplanets are predicted to remain with
 stellar compositions, at least if they have solar composition. \par
 
In paper I we present the masses of heavy elements that can be accreted by protoplanets
, with solar composition and   masses between 3 and 10 M$_J$, at large radial distance. 
 The exact amount of solids that can be captured depends on the disk properties, the planetary location, and the planetesimals' sizes and velocities. 
 For conditions similar to the ones expected for the HR 8799 planetary system, we find that
 a 10 M$_J$ planet could accrete between 0 and 90 M$_{\oplus}$ of heavy elements, if
 formed between 24 and 68 AU, respectively. Planets of 7 and 5 M$_J$ can capture between 0
 and 46, and 0 and 23 M$_{\oplus}$ of solids, at radial distances between 24 and 68 AU,
 respectively.  A planet of 3  M$_J$, at radial distances between 24 and 68 AU, is found
 to accrete up to 37.5 M$_{\oplus}$  of heavy elements. Although in some cases a non-negligible mass of heavy elements can be accreted, the overall planetary composition remains close to stellar. 
But 
the longer evolution of metal-rich protoplanets, as just described, 
 could lead to  much more significant accretion of solids 
at wide orbits.  On the other hand, metal-rich objects will be more vulnerable to 
 tidal disruption and therefore may not survive.  Giant protoplanets with low metallicities
 are likely to stay metal poor.  \par

The picture changes significantly when grain coagulation and sedimentation are considered
(section 3). In that case we find that the pre-collapse timescales can be orders of
 magnitude shorter. For example, for the 5 M$_J$ solar case the time scale is reduced 
from 26,000 to 1600 years, and for the metal-rich case the reduction is from 75,000
to 700 years. The characteristic pre-collapse timescale is found to be of the order of 
one thousand years.  The exact numbers change with planetary mass and metallicity, but 
in general it is found that when grain growth and settling are included there is much less
 sensitively to the planetary mass and the assumed metallicity. Another result which
 differs considerably from the one presented in section 2 is that metal-rich planets
 actually have shorter contraction timescales  than metal-poor or solar-composition 
planets, due to their efficient grain growth resulting  from their original large
 number densities, although the difference in timescale is not always large. The results
 from section 3 suggest that there is not much opportunity to capture solids for a very
 large range of planetary masses and compositions, because of the sharply reduced time
scales. 
To test the possibility of core formation we follow the grain settling all the way to the
 planetary center. We find that in all cases 
 although there is a substantial settling in the outer layers that leads to opacity reduction, grains cannot settle to the 
 innermost region and form cores. 
In this scenario we find that the planetary objects are likely to remain with stellar abundance and contain no cores. \par

Recently, Boley and Durisen (2010) have found that 
gravitationally unstable clumps formed in spiral arms could have enriched initial concentrations of heavy elements as well as possible concentration of solid material toward the center of the clump. 
This configuration might favor core formation in protoplanets at early stages of the evolution. Although in this scenario cores could be formed, more detailed calculations are needed before a robust conclusion can be reached.  

\section{Summary}

We investigate the effect of opacity on the early evolution of giant gaseous protoplanets 
which have formed by gravitational instability in the outer regions of protoplanetary
disks. 
First, we model the evolution of 3, 5, and 7 M$_J$ protoplanets with different
 metallicities simply by scaling their grain opacity with metallicity. In that case we
 find that the pre-collapse timescale is proportional to planetary metallicity; metal-poor 
protoplanets have the short contraction timescales while metal-rich protoplanet are found 
to contract over significantly longer timescales.  We note that the shorter pre-collapse
 stages of metal-poor protoplanets can support their survival. Metal-rich protoplanets on 
the other hand, spend longer times in the extended phase and therefore are predicted to be
 more vulnerable to destruction. However, if they do manage to survive, they have better
 opportunity to accrete solid planetesimals and form heavy-element cores.  \par

We next investigate the effect of opacity reduction on the pre-collapse evolution due to
 grain coagulation and sedimentation. We find that including these processes leads to
 significantly shorter evolutionary timescales for all the planetary masses and 
metallicities considered. The pre-collapse timescale is found to be of the order of one
thousand years for masses between 3 and 7 M$_J$, and is relatively insensitive to planetary
 composition.  We find that in this scenario the pre-collapse evolution of  a metal-rich
 protoplanet is actually shorter than that of a metal-poor planet, due to very efficient
 opacity reduction caused by the larger amounts of grains initially present in the
 atmosphere, that lead to rapid grain growth and settling.  The considerably shorter
 pre-collapse stages found in this scenario may reduce the risk of clump disruption, and
 support their survival. Since clumps accrete gas (and solids) most efficiently during their early evolution, 
 shortening the pre-collapse stage would also limit the growth of the  newly-formed protoplanets and would lead to smaller final masses.  
On the other hand, the short pre-collapse evolution would 
lead to negligible enrichment with heavy elements and will suppress the formation of cores in these objects. 
The results for the evolutionary timescales for the two cases we consider are summarized in Table 1.  \par

The presented work cannot provide strong predictions due to the complex nature of the
 problem. Here we show, in a simple one-dimensional approximation, that adding grain 
coagulation and settling leads to very different results from the ones obtained when 
these processes are not included. Although our results suggest that massive planets formed by gravitational instabilities are unlikely to be significantly metal enriched relative to the star, observations of giant planets with stellar compositions at the surface should not be taken as evidence for formation by gravitational instability. Massive giant planets formed by core accretion could also have a nearly stellar overall composition. In the standard core accretion model the typical core (i.e. heavy elements) masses do not exceed several Earth masses, while the rest of the planetary mass consists of residual nebular gas (Hubickyj et al., 2005; Movshovitz et al., 2010). The exact composition of the gas would depend on disk properties, orbital location, formation timescale, etc. If the composition of the residual gas is stellar, or somewhat depleted in heavy elements compared to the star, the total abundance of the forming giant planet would be about stellar. As a result, giant planets with nearly stellar compositions could be explained by both core accretion and gravitational instability. \par

Naturally, there are more processes that could affect
 the early evolution and survival of giant protoplanets. For example, ongoing accretion of nebular gas with small dust grains would increase the opacity at the outer layers resulting in less rapid contraction. In addition, accretion would affect the planetary luminosity and therefore its contraction time. Planetary rotation could also lead to slower contraction, however, the real effect of angular momentum has to be calculated in detail before a definite conclusion can be reached. 
Our calculation does not assume a specific location for the formation of the planet, except that it has to be outside 20 AU for tidal stability. In actual disks protoplanets could migrate inward or outward, a process which could also affect the planets' survival. \par
 
In this work we assume that all solids are in the form of small grains. However, if a substantial fraction of the solid material were in the form of planetesimals the opacity and the timescale for evolution would be affected. In the case of no settling (case I) the evolution times would be shorter than the ones given in table 1 due to lower opacity.  In case II, the evolution timescale would be somewhat longer due to slower grain growth. It is unclear whether appreciable planetesimal formation had occurred at early stages of clump formation, especially at the large radial distances which we consider ($>$ 20 AU). 
We have now added one missing piece to the puzzle, and hope that future investigations including additional effects will provide a clearer picture of early planetary evolution.

\section*{Acknowledgments}
P. B.  acknowledges support from NASA Origins grant
   NNX08AH82G and NSF grant AST-0908807. R. H. acknowledges
 support from NASA through the Southwest Research Institute. 

\section*{References} 
%\small{
\noindent Alexander, D., Ferguson, J., 1994. Low-temperature Rosseland opacities. 
Astrophys. J. 437, 879--891.\\
\noindent Bodenheimer, P., Grossman, A. S., Decampli, W. M., Marcy, G., Pollack,
 J. B., 1980. Calculations of the evolution of the giant planets. Icarus 41, 293--308.\\
Boley, A. C., 2009. The two modes of giant planet formation. Astrophys. J. 
695, L53--56. \\
Boley, A. C., Hayfield, T., Mayer, L., Durisen, R. H., 2010. 
Clumps in the outer disk by disk instability: Why they are initially gas giants and the legacy of disruption. Icarus, 207, 509--516.\\
Boley, A. C and Durisen, R. H., 2010. Enrichment and Differentiation in Gas Giants During Birth by Disk Instability. arXiv:1005.2624\\
Boss, A. P., 1997. Giant planet formation by gravitational instability. Science 276,
1836--1839.\\
Boss, A. P., 2002. Evolution of the solar nebula. V. Disk instabilities with varied
 thermodynamics.  Astrophys. J. 576, 462--472. \\
Boss, A. P., 2008. Flux-limited diffusion approximation models of giant planet
formation. Astrophys. J. 677, 607-615. \\
Cai, K., Durisen, R. H., Michael, S., Boley, A. C., Mej'a, A. C., Pickett, M. K., D'Alessio, P. The Effects of Metallicity and 
Grain Size on Gravitational Instabilities in Protoplanetary Disks. 2006a, ApJ, 636, L149--L152.\\
Cai, K., Durisen, R. H., Michael, S., Boley, A. C., Mej'a, A. C., Pickett, M. K., D'Alessio, P. Erratum: ``The Effects of Metallicity and Grain Size on 
Gravitational Instabilities in Protoplanetary Disks'', 2006b, ApJ, 642, L173--L173.\\
Cai, K., Pickett, M. K., Durisen, R. H., Milne, A. M., 2009.
 Giant planet formation by disk instability: A comparison simulation with an 
improved radiative scheme.  Astrophys. J. Letters, submitted. arXiv:0907.4213 \\
Cameron, A. G. W., 1978. Physics of the primitive solar nebula and of giant
 gaseous protoplanets. In: Gehrels, T. (Ed.) Protostars and planets: Studies of star
 formation and of the origin of the solar system.  University of Arizona Press, 
Tucson, pp. 453--487. \\
Decampli, W. M., Cameron, A. G. W., 1979. Structure and evolution of isolated giant
 gaseous protoplanets. Icarus 38, 367--391.\\
Dodson-Robinson, S., Veras, D., Ford, E., Beichman, C. A., 2009. The formation
mechanism of gas giants on wide orbits. Astrophys. J. 707, 79-88. \\
Durisen, R. H., Boss, A. P., Mayer, L., Nelson, A. F., Quinn, T., Rice, W. K. M. , 2007.
Gravitational instabilities in protoplanetary disks and implications for giant 
planet formation. In Reipurth, B. et al. (Eds.), Protostars and Planets V, 
Univ. of Arizona Press, Tucson,  pp. 607-622.\\
Gammie, C. F., 2001. Non-linear outcome of gravitational instability in cooling, gaseous
disks. Astrophys. J. 553, 174-183. \\
Helled, R., Podolak, M., Kovetz, A., 2006. Planetesimal capture in the disk 
instability model. Icarus 185, 64--71.\\
Helled, R., Podolak, M., Kovetz, A., 2008. Grain sedimentation in a giant gaseous
 protoplanet. Icarus 195, 863--870. \\
 %Helled, R., Schubert, G., 2009. Heavy-element enrichment of a Jupiter-mass 
%protoplanet as a function of orbital location.  Astrophys. J. 697, 1256--1262. \\
 Helled, R., Bodenheimer, P., 2010. Metallicity of the massive protoplanets around 
HR 8799 if formed by gravitational instability. Icarus, 207, 503--508 (paper I). \\
Hubickyj, O., Bodenheimer, P., Lissauer, J. J., 2005. 
 Accretion of the gaseous envelope of Jupiter around a 5--10 Earth-mass core.
 Icarus 179,  415--431.\\
Kalas, P. et al.,  2008.  Optical imaging of an exosolar planet 25 light 
years from Earth.  Science 322, 1345--47. \\
Marois, C., Macintosh, B., Barman, T.,  Zuckerman, B., Song, I., Patience, J.,
 Lafreniere, D., Doyon, R., 2008. Direct imaging of multiple planets orbiting the
 star HR 8799. Science 322, 1348--1352.\\
Mayer, L., Lufkin, G., Quinn, T.,   Wadsley, J., 2007. Fragmentation of 
gravitationally unstable gaseous protoplanetary disks with radiative transfer.
 Astrophys. J. 661, L77--L80.\\
Meru, F., Bate, M. R., 2010. Exploring the conditions required to form giant
planets via gravitational instability in massive protoplanetary disks. 
arXiv:1004.3766.\\
Movshovitz, N.,  Podolak, M., 2008. The opacity of grains in protoplanetary atmospheres, Icarus, 194, 368--378.\\
Movshovitz, N., Bodenheimer P., Podolak, M., Lissauer, J. J., 2010.  
Formation of Jupiter with opacities based on detailed grain 
physics, Icarus, in press. arXiv:1005.3875 \\
Ormel, C. W.,  Cuzzi, J. N., 2007. Closed-form expressions for particle 
relative velocities induced by turbulence. Astron. Astrophys, 466, 413--420. \\ 
Podolak, M., 2003. The contribution of small grains to the opacity of 
protoplanetary atmospheres. Icarus 73, 163-179.\\
Pollack, J. B., Hubickyj, O., Bodenheimer, P., Lissauer, J. J., Podolak, M.,
 Greenzweig, Y., 1996. Formation of the giant planets by concurrent accretion of
 solids and gas. Icarus 124, 62--85.\\
Pollack, J., McKay, C., Christofferson, B., 1985. A calculation of the Rosseland
mean opacity of dust grains in primordial Solar System nebulae.  Icarus 64, 
471-492.\\
Rafikov, R. R., 2007. Convective cooling and fragmentation of gravitationally 
unstable disks. Astrophys. J. 662, 642--650.\\ 
Rafikov, R. R., 2009. Properties of gravitoturbulent accretion disks. Astrophys. J. 704, 281--291.\\
van de Hulst, H. C., 1957. Light scattering by small particles. John Wiley \& Sons, 
New York.\\

\newpage
\section{Appendix 1: A simple approximation for grain growth timescale}
The coagulation time can be estimated by dividing the mean free path for grain-grain collisions by the random (thermal) speed of the grains. Taking $X$ to be the
mass fraction of dust in the gas (changing with metallicity), the total mass of grains within the body is given by $M_{grain}=XM_{gas}$, where $M_{gas}$ is the mass of the gas.
Assuming that all the grains within the protoplanet have a similar size, $a_0$, the number density of grains can be written as
\begin{equation}
n_{grain} = {{3X \rho} \over 4 \pi \rho_{grain} a_0^3}
\end{equation} 
with $\rho$ and $\rho_{grain}$ being the density of the planetary envelope and the grain, respectively. 

The cross section of the grains is given by $\sigma=4\pi a_0^2$, and the
mean free path between collisions is
\begin{equation}
 \lambda = {1 \over {\sqrt 2 {n \sigma}}} =  {{\rho_{grain}a_0} \over {3 \sqrt{2}X
 \rho}}.
\end{equation}
The velocity of the grains is taken to be the thermal velocity $v_{th}$ is given by
\begin{equation}
v_{th} = \sqrt{8kT \over \pi m}
\end{equation}
where $T$ is the temperature of the atmosphere, $k$ is Boltzmann's constant, and $m$ being the
grain's mass.\\
Finally, the time between collisions is given by
\begin{equation}
\ \tau_{coag} = {{\lambda \over v_{th}}}= {{\rho_{grain}a_0}
\over {3 \sqrt{2}X \rho}} {{\sqrt{\pi m_{grain}\over 8kT }}}=
{{\pi \over 6X \rho}{\sqrt{\rho_{grain}^3 a_0^5 \over 3kT}}}.
\end{equation}
Assuming that the atmospheric properties are about the same for protoplanets with different
 metallicities, one can see from equation (5) that the coagulation time is inversely 
proportional to $X$, the dust-to-gas ratio. As a result, grain growth (followed  by
 grain settling) is more efficient in metal-rich protoplanets.
This is only a rough estimate for the coagulation time,  since it does not include growth
 in which a grain with higher sedimentation velocity overtakes grains with smaller
 sedimentation speeds, a mechanism that can be very efficient especially for small grain 
sizes. Once a slightly bigger grain is formed, it can immediately sweep up the small
 grains in the background. The approximation above is valid only for a radiative envelope, 
however in the models considered here the opacity in the outer radiative envelope
essentially determines the time scale of evolution. The coagulation  time in the
outer layers of these models at the early stages of evolution is only a few years
for $a_0 = 10^{-5}$ cm.

\clearpage

\begin{table}[h!]
\begin{center}
{\renewcommand{\arraystretch}{0.6}
\vskip 8pt
\begin{tabular}{lc c c c|}
\hline
& & Planetary Composition&\\
Planetary Mass & solar/3 & solar &3$\times$solar \\
\hline
3 M$_J$ - case I & 29,075 yrs & 78,189 yrs & 223,705 yrs \\
3 M$_J$ - case II &  2,337 yrs  & 1,667 yrs & 1,062 yrs\\
\hline
5 M$_J$ - case I & 9,119 yrs  & 26,160 yrs & 75,379 yrs\\
5 M$_J$ - case II & 1,707 yrs &  1,587 yrs & 720 yrs\\
\hline
7 M$_J$ - case I & 3,880 yrs  & 11,770 yrs & 34,840 yrs\\
7 M$_J$ - case II & 1,336 yrs  & 1,304 yrs & 474 yrs\\
\hline
\end{tabular} 
}
\caption{\label{data} 
Pre-collapse evolution timescales for the two different cases considered: Case I:  the opacity is scaled with planetary metallicity and is constant with time. Case II: the opacity can be reduced with time as a
result of grain growth and settling.
}
\end{center} 
\end{table}

%\clearpage

\begin{figure}[h!]
   \centering
   \includegraphics[width=4.0in]{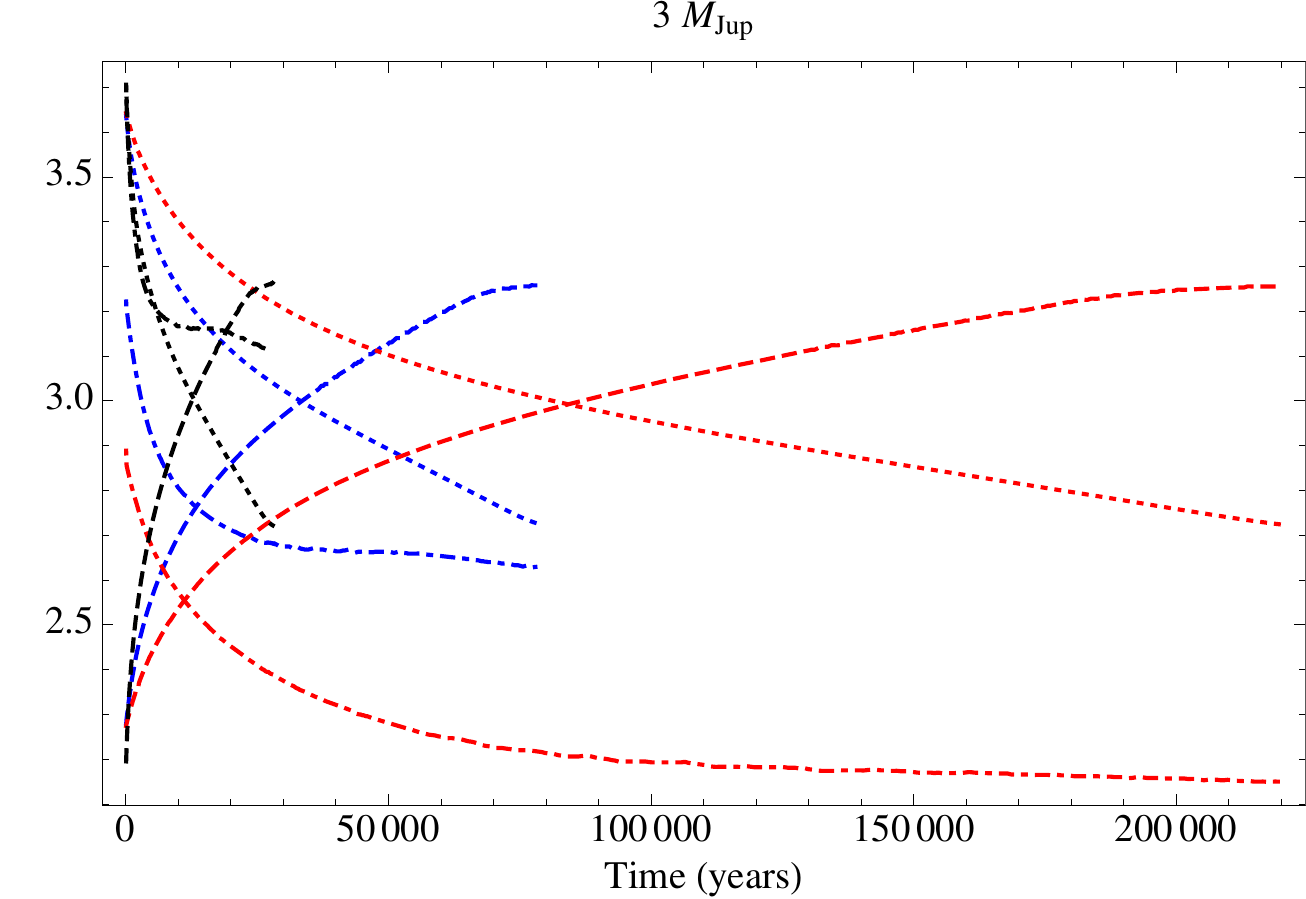}
    \caption[err]{Physical properties as a function of time for 3 Jupiter mass protoplanets with solar (blue), three times solar (red) and solar/3 (black) compositions. The dashed, dotted, and dot-dashed curves represent Log(T$_\text{c}$) [K],  Log(R/R$_{\text{Jup}}$) [cm], and Log(L/10$^{26}$) [erg s$^{-1}$], respectively.}
\end{figure}

\begin{figure}[h!]
   \centering
   \includegraphics[width=4.0in]{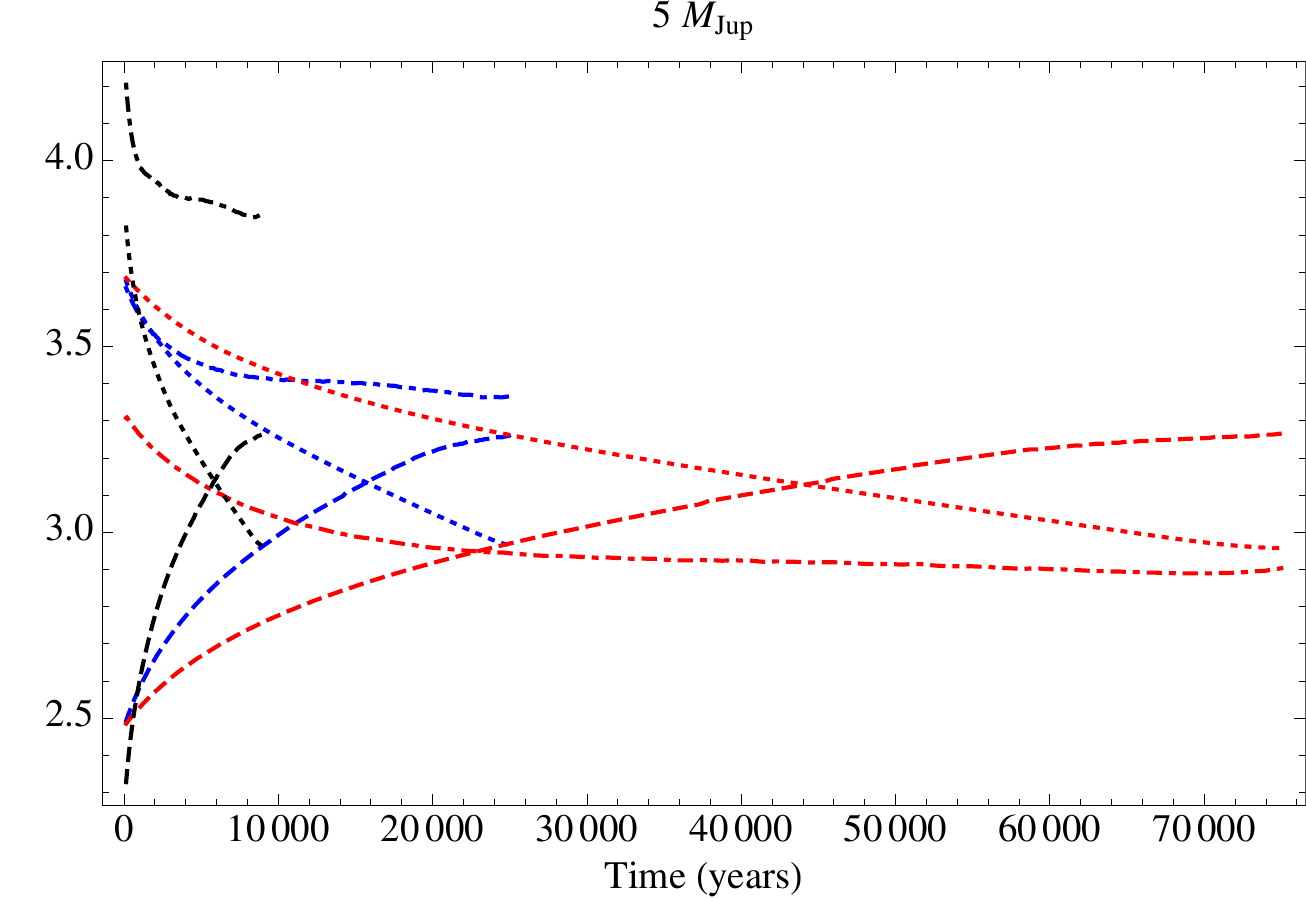}
    \caption[err]{Physical properties as a function of time for 5 Jupiter mass protoplanets with solar (blue), three times solar (red) and solar/3 (black)  compositions. The dashed, dotted, and dot-dashed curves represent Log(T$_\text{c}$) [K],  Log(R/R$_{\text{Jup}}$) [cm], and Log(L/10$^{26}$) [erg s$^{-1}$], respectively.}
\end{figure}

\begin{figure}[h!]
   \centering
   \includegraphics[width=4.0in]{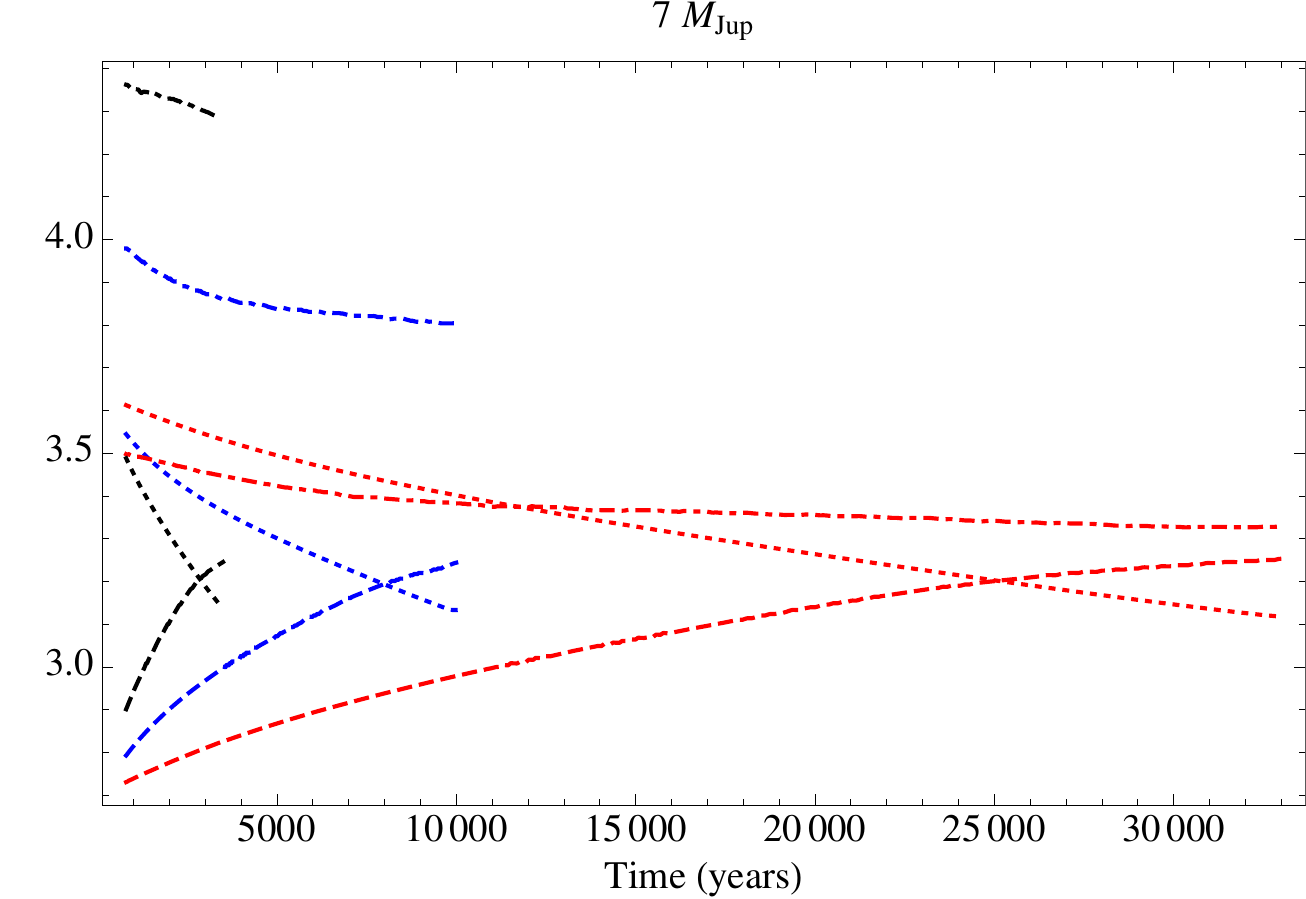}
    \caption[err]{Physical properties as a function of time for 7 Jupiter mass protoplanets with solar (blue), three times solar (red) and solar/3 (black) compositions. The dashed, dotted, and dot-dashed curves represent Log(T$_\text{c}$) [K],  Log(R/R$_{\text{Jup}}$) [cm], and Log(L/10$^{26}$) [erg s$^{-1}$], respectively.}
\end{figure}

%\newpage
\begin{figure}[h!]
   \centering
   \includegraphics[width=6.0in]{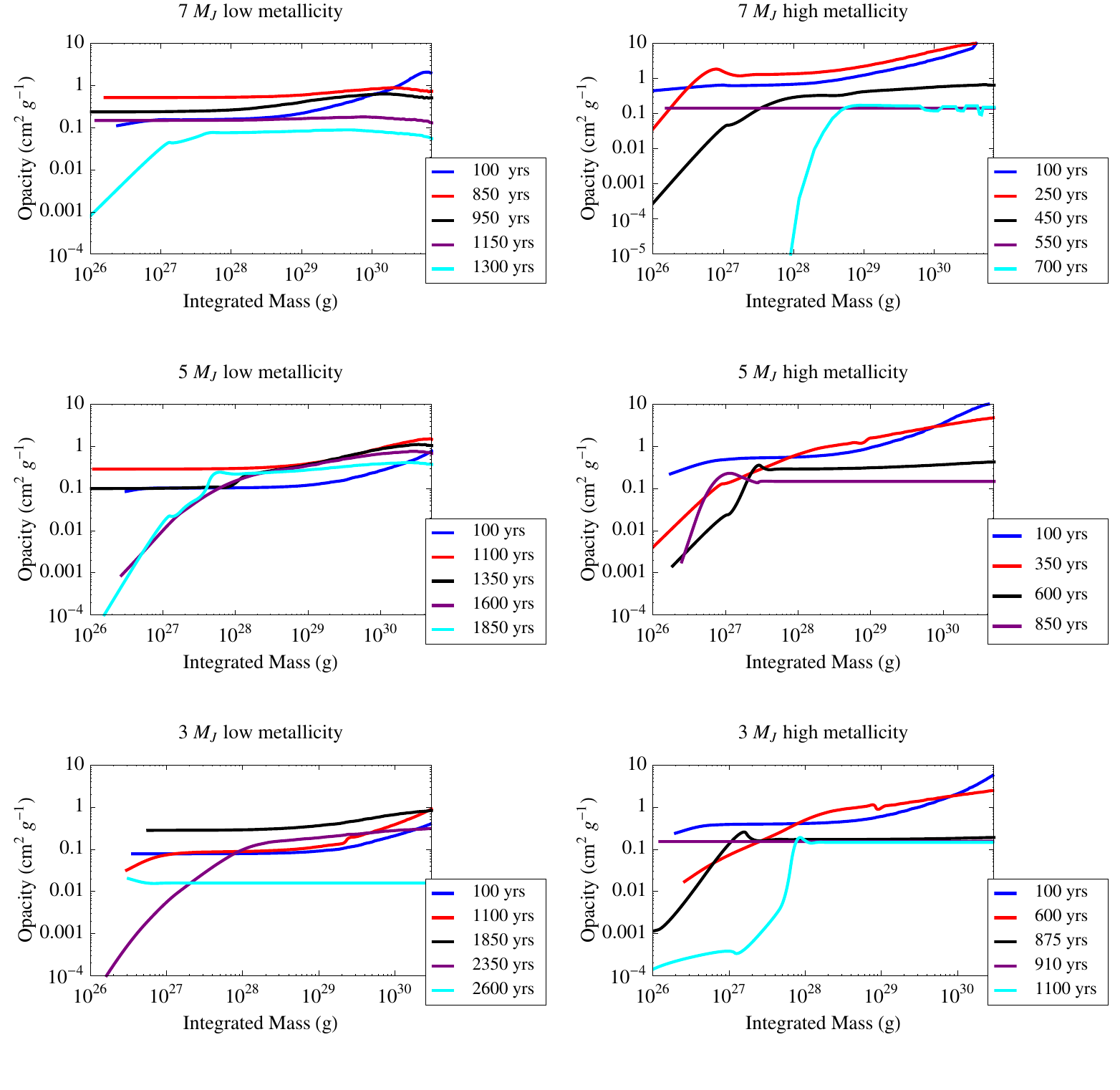}
    \caption[err]{
Rosseland mean opacity as a function of depth in the protoplanetary
envelope at various times when grain coagulation and sedimentation are included. The
left-hand panels correspond to a metal abundance of 1/3 solar; the right-hand panels
to 3 times solar.  The abscissa refers to the mass integrated inward from the surface.}
\end{figure}

\newpage
\begin{figure}[h!]
   \centering
   \includegraphics[width=3.35in]{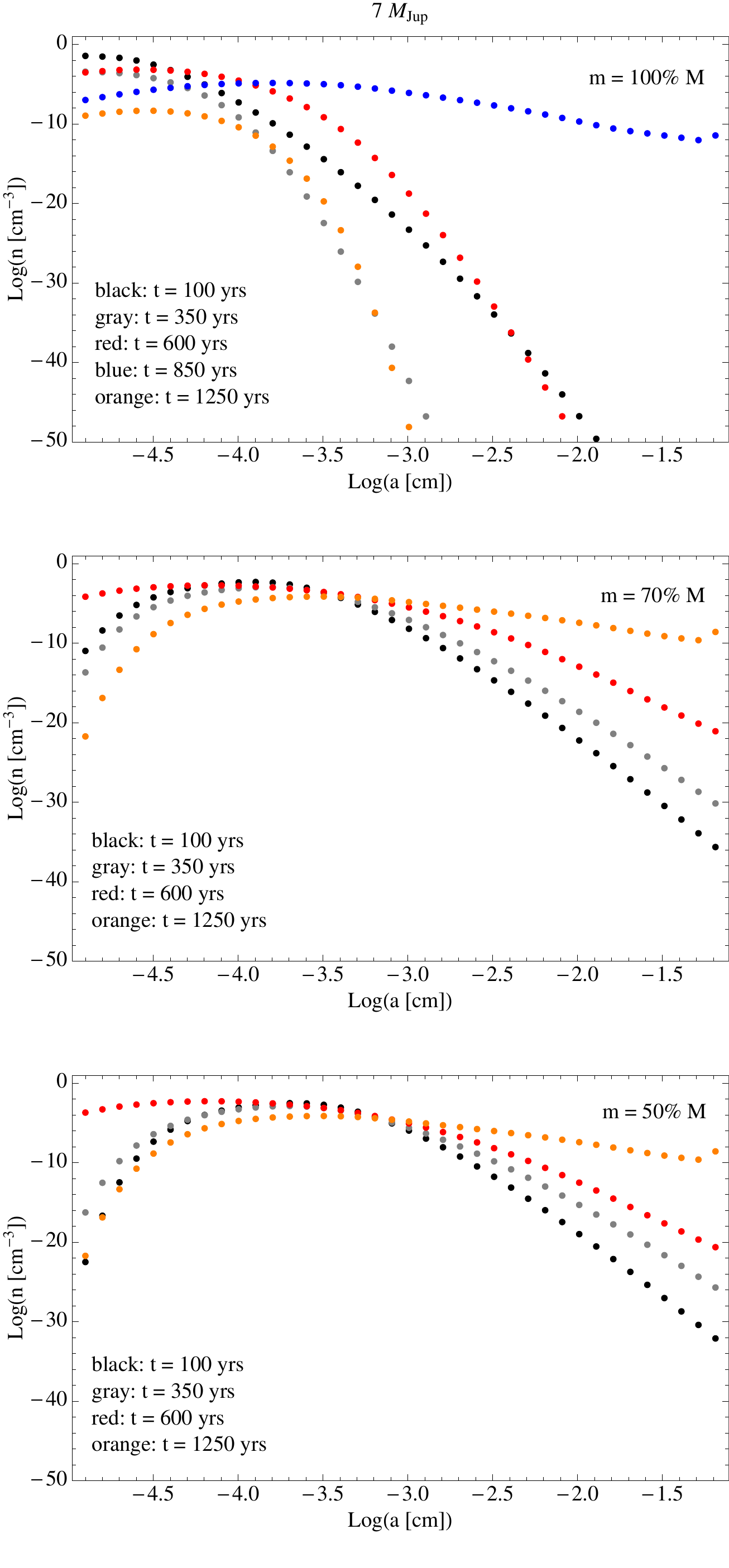}
    \caption[err]{
Grain number density 'n' vs.~grain size 'a' for metal-poor 7 M$_J$ planet at different times. The three different figures correspond to three different layers  in the envelope defined by the integrated mass outward from the planetary center. The figures are presented for integrated masses of 100 (top), 70 (middle), and 50 (bottom) percent of the total planetary mass. Different colors correspond to different times.}
\end{figure}

\newpage
\begin{figure}[h!]
   \centering
   \includegraphics[width=4.0in]{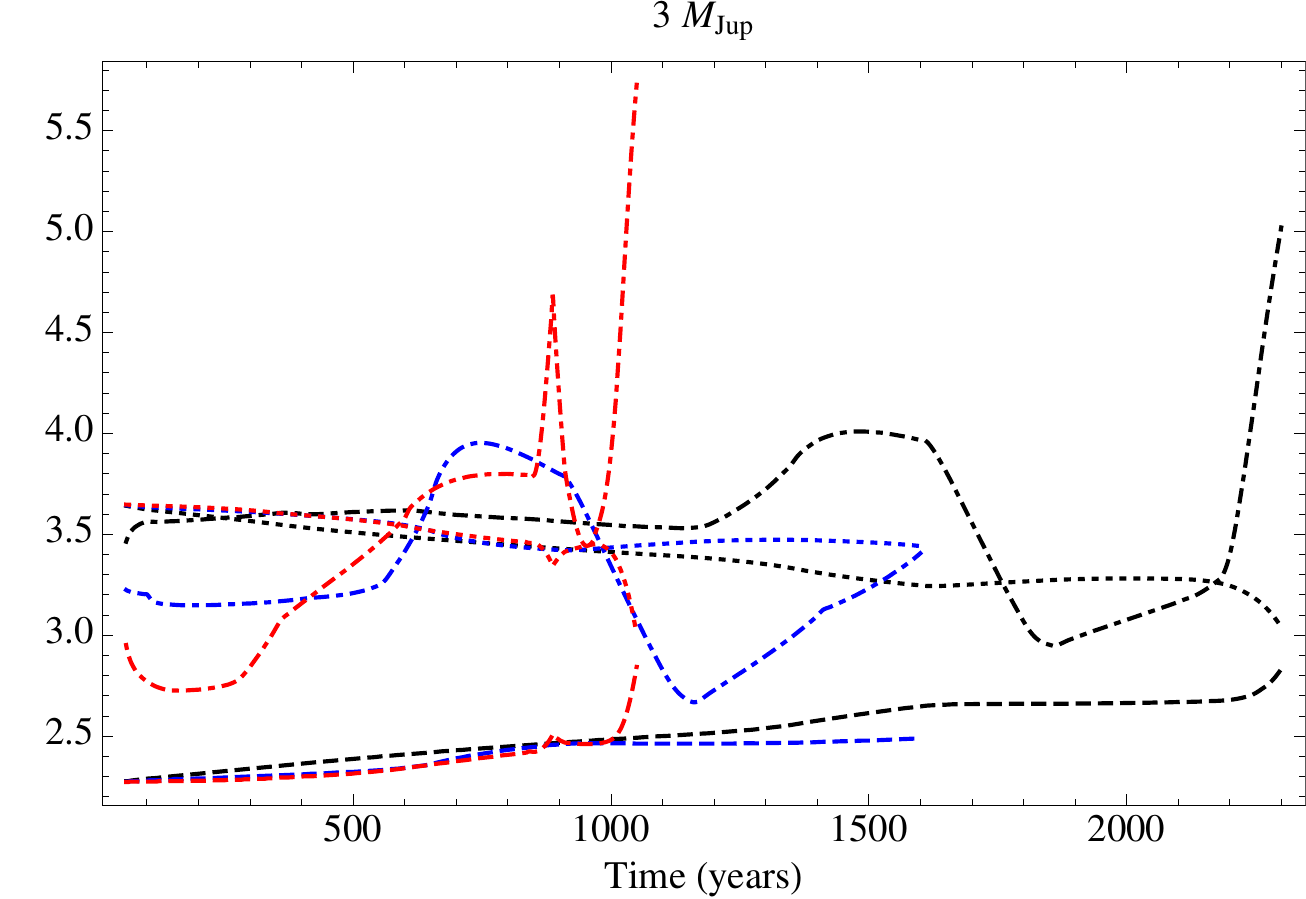}
    \caption[err]{Physical properties as a function of time when grain growth and settling are
 included for 3 Jupiter mass protoplanets with solar (blue), three times solar (red) and solar/3 (black) compositions. The dashed, dotted, and dot-dashed curves represent Log(T$_\text{c}$) [K],  Log(R/R$_{\text{Jup}}$) [cm], and Log(L/10$^{26}$) [erg s$^{-1}$], respectively.}
\end{figure}

\begin{figure}[h!]
   \centering
   \includegraphics[width=4.0in]{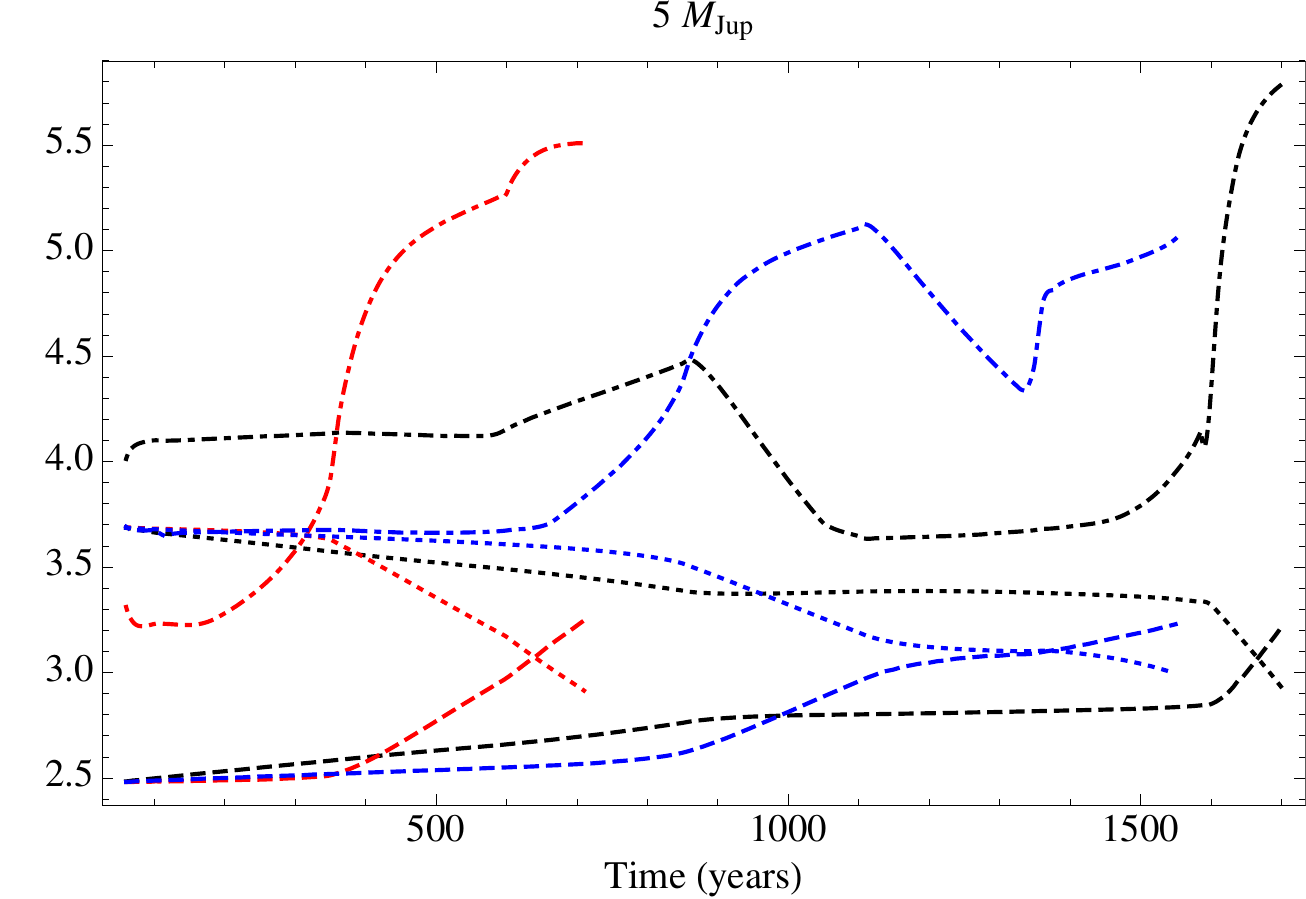}
    \caption[err]{Physical properties as a function of time when grain growth and settling are
included for 5 Jupiter mass protoplanets with solar (blue), three times solar (red) and solar/3 (black) compositions. The dashed, dotted, and dot-dashed curves represent Log(T$_\text{c}$) [K],  Log(R/R$_{\text{Jup}}$) [cm], and Log(L/10$^{26}$) [erg s$^{-1}$], respectively.}
\end{figure}

\begin{figure}[h!]
   \centering
   \includegraphics[width=4.0in]{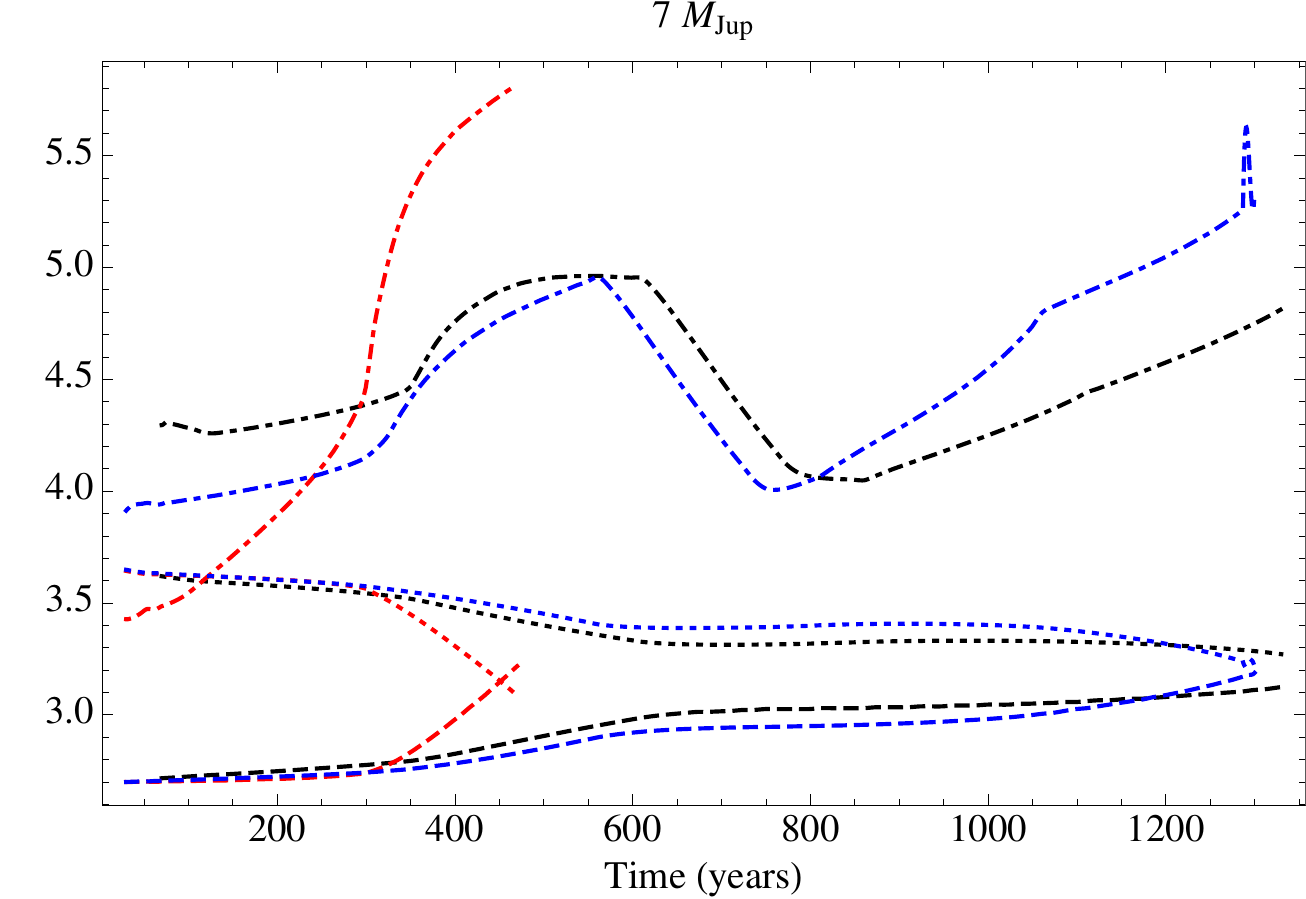}
    \caption[err]{Physical properties as a function of time when grain growth and settling are
 included for 7 Jupiter mass protoplanets with solar (blue), three times solar (red) and solar/3 (black) compositions. The dashed, dotted, and dot-dashed curves represent Log(T$_\text{c}$) [K],  Log(R/R$_{\text{Jup}}$) [cm], and Log(L/10$^{26}$) [erg s$^{-1}$], respectively.}
\end{figure}

\begin{figure}[h!]
   \centering
   \includegraphics[width=4.0in]{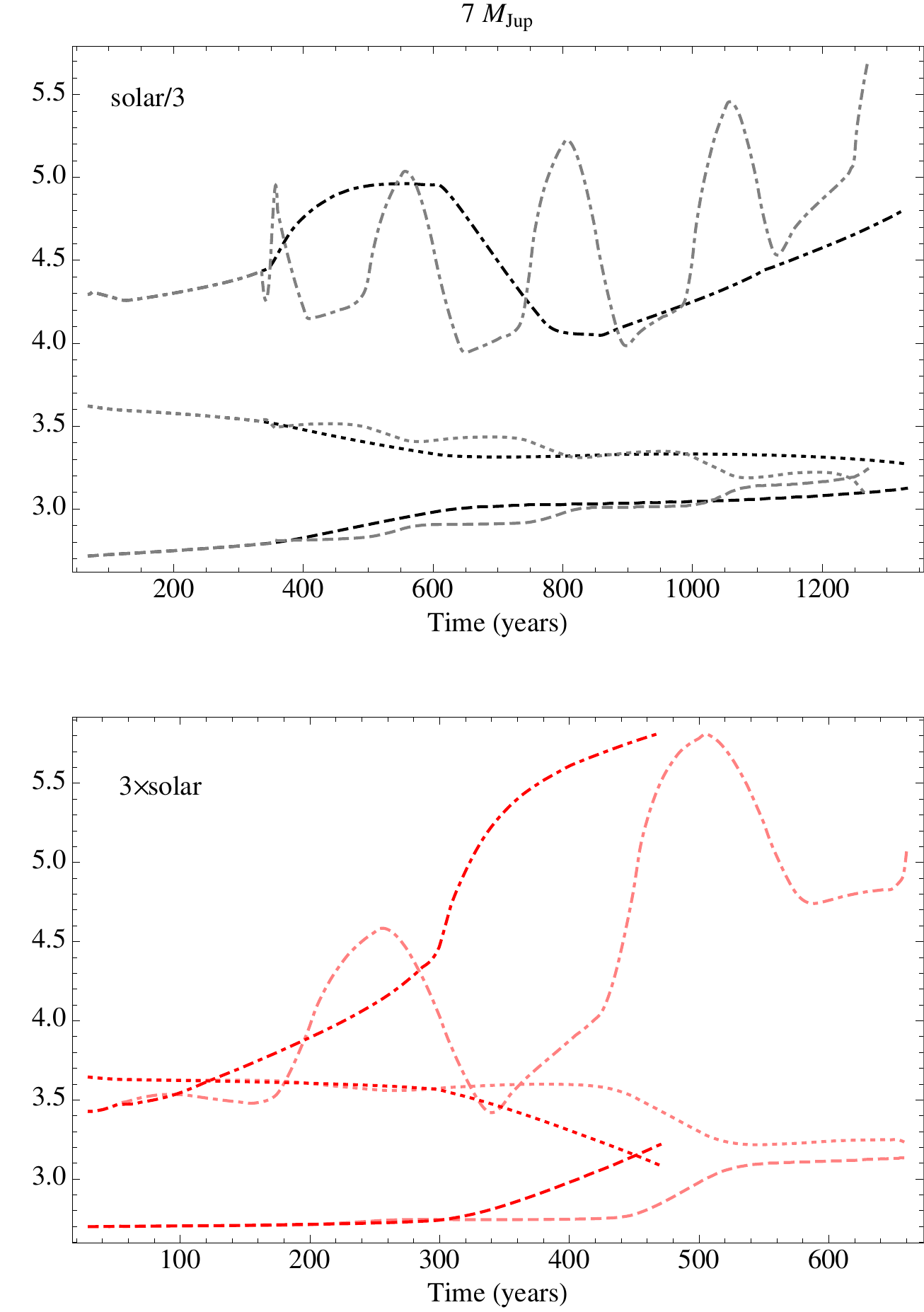}
    \caption[err]{Physical properties as a function of time when grain growth and settling are
 included for 7 M$_J$ protoplanets with solar/3 (top) and 3$\times$solar (bottom) compositions with long (black, red; 250 years) and short (gray, pink; 50 years) time intervals between opacity recalculations. Again, the dashed, dotted, and dot-dashed curves represent Log(T$_\text{c}$) [K],  Log(R/R$_{\text{Jup}}$) [cm], and Log(L/10$^{26}$) [erg s$^{-1}$], respectively.}
\end{figure}
\end{document}